\documentclass[aps,prl,twocolumn,superscriptaddress,showpacs,longbibliography]{revtex4-1}
\usepackage{amsmath,amssymb,graphicx,subfigure}

\usepackage[utf8]{inputenc}
\usepackage[T1]{fontenc}
\usepackage{xcolor}
\usepackage[normalem]{ulem}

\IfFileExists{newtxtext.sty}
   {\usepackage{newtxtext,newtxmath}}
   {\IfFileExists{stix.sty}
      {\usepackage{stix}}
      {\IfFileExists{mathptmx.sty}
      {\usepackage{mathptmx}}{} } }

\usepackage{textcomp}

\usepackage{bm}

\IfFileExists{siunitx.sty}{\usepackage{booktabs,siunitx}}{}

\pdfoutput=1
\usepackage{color}
\definecolor{LinkColor}{rgb}{0.256,0.439,0.588}
\usepackage{hyperref}
\hypersetup{
   colorlinks=true,
   citecolor=LinkColor,
   linkcolor=LinkColor,
   urlcolor=LinkColor
}

\newcommand{\bra}[1]{\langle#1\rvert}
\newcommand{\ket}[1]{\lvert#1\rangle}

\newcommand{\be}{\begin{equation}}
\newcommand{\ee}{\end{equation}}
\newcommand{\bea}{\begin{eqnarray}}
\newcommand{\eea}{\end{eqnarray}}

\usepackage{pifont}

\begin{document}

\title{Spontaneous fractional Chern insulators in transition metal dichalcogenides Moir{\'e} superlattices}
\author{Heqiu Li}
\affiliation{Department of Physics, University of Michigan, Ann Arbor, Michigan 48109, USA}
\author{Umesh Kumar}
\affiliation{Theoretical Division, T-4, Los Alamos National Laboratory, Los Alamos, New Mexico 87545, USA}
\author{Kai Sun}
\affiliation{Department of Physics, University of Michigan, Ann Arbor, Michigan 48109, USA}
\author{Shi-Zeng Lin}
\affiliation{Theoretical Division, T-4 and CNLS, Los Alamos National Laboratory, Los Alamos, New Mexico 87545, USA}

\date{\today}

\begin{abstract}
Moir{\'e} superlattice realized in two-dimensional heterostructures offers an exciting platform to access strongly-correlated electronic states. In this work, we study transition metal dichalcogenides (TMD) Moir{\'e} superlattices with time-reversal symmetry and nontrivial spin{/valley}-Chern numbers.
Utilizing realistic material parameters and the method of exact diagonalization, we find that at certain twisting angle and fractional filling, gapped fractional topological states, i.e., fractional Chern insulators, are naturally {stabilized} by simply introducing the Coulomb repulsion. In contrast to fractional quantum Hall systems,  where the time-reversal symmetry has to be broken explicitly, these fractional states break the time-reversal symmetry spontaneously. {We show that the Chern number contrasting in the opposite valleys imposes a strong constraint on the nature of fractional Chern insulator and the associated low energy excitations.}
\end{abstract}

\maketitle

\emph{Introduction.}--- When two layers of two-dimensional materials are placed atop of each other with slight misalignment, it creates a superlattice with periodicity much larger than the atomic lattice parameter. Because of the large lattice periodicity, one can fill or empty the entire band by electrode gating. This Moir{\'e} superlattice provides a tunable platform to control the electronic band structure \cite{PhysRevLett.99.256802, bistritzer_moire_2011}, and therefore enables access to a plethora of interesting quantum states. Because the band width in these systems can be tuned to be extremely narrow 
\cite{bistritzer_moire_2011}, these Moir{\'e} superlattices open up a new pathway to stabilize various strongly-correlated phases such as superconductivity and correlated insulators \cite{cao_correlated_2018,cao_unconventional_2018,lu_superconductors_2019,Yankowitz1059,kerelsky2019maximized,cao2019strange,polshyn2019large,xie2019spectroscopic,jiang2019charge,choi2019electronic,zondiner_cascade_2020,wong_cascade_2020,nuckolls_strongly_2020,he_symmetry_2020,liu_tunable_2020,regan_mott_2020,wang_correlated_2020,PhysRevLett.124.097601,wu2019collective,YingLinPRL2020,Padhi2018,Padhi2020,Padhi2019,Stefanidis2020,PhysRevLett.124.166601}. Furthermore, such electronic band structure can also be topologically nontrivial, e.g., with a nonzero integer Chern number \cite{PhysRevB.99.075127,PhysRevLett.124.187601,Sharpe605,youngscience,chen2019tunable}. Combined with their strong coupling nature, such Moir{\'e} superlattices offer a promising route to realize the long-sought fractionalized topological order \cite{PhysRevResearch.2.023237,PhysRevResearch.2.023238,PhysRevLett.124.106803,liu_gate-tunable_2020,wilhelm_interplay_2020,Sohal2018,Sohal2020}.

Recently, gapped electronic states at various fractional fillings (e.g., $1/3$) were observed in transition metal dichalcogenide (TMD) Moir\'{e} superlattices, e.g., $\mathrm{WSe_2/WS_2}$ \cite{regan_mott_2020,xu_correlated_2020,jin_stripe_2020,zhou_signatures_2020,huang_correlated_2020}. In general, gapped electronic states at fractional filling may have two origins: (a) charge order that spontaneously breaks the translational symmetry and (b) fractional topological order, e.g., fractional Chern insulators (FCI)~\cite{PhysRevLett.106.236802,PhysRevLett.106.236803,PhysRevLett.106.236804,Regnault2011,Sheng2011,parameswaran_fractional_2013,bergholtz_topological_2013,PhysRevB.85.075116}. In these TMD Moir\'{e} superlattices, the observed gapped states were interpreted as Wigner crystals of electrons,  because the underlying single-particle bands are topologically trivial~\cite{PhysRevLett.121.026402}.

Encouraged by such exciting experimental progress, here we explore the feasibility of the second category in TMD Moir\'{e} superlattices. In particular, we focus on systems like $\mathrm{MoTe_2}$, which may host topologically nontrivial bands with non-zero spin/valley-Chern numbers~\cite{Wu2019}.
In contrast to a partially filled Chern band~\cite{PhysRevLett.106.236802,PhysRevLett.106.236803,PhysRevLett.106.236804,Regnault2011,Sheng2011,parameswaran_fractional_2013,bergholtz_topological_2013,PhysRevB.85.075116}, because these systems preserve the time-reversal symmetry, two types of fractional states are in principle allowed (a) time-reversal invariant fractional topological insulators~\cite{PhysRevLett.103.196803}  and (b) FCIs via spontaneously breaking the time-reversal symmetry. The key focus of this study is whether Coulomb repulsion could stabilize some of these fractional states in TMD Moir\'{e} superlattices.

In this work,  we show that by simply increasing the Coulomb interaction strength in such TMD Moir\'{e} superlattices, the system undergoes a quantum phase transition that spontaneously breaks the time-reversal symmetry by polarizing electrons into one of the two valleys. Further increase of Coulomb interaction will trigger a second quantum phase transition, and stabilize a FCI at a fractional filling.
For excitations, our numerical studies observe both (intravalley) fractional excitations from the fractional topological order and (intervalley) valley-wave excitations from the spontaneous symmetry breaking. {We argue that the symmetry breaking state and low-energy excitations are constrained by the valley contrasting Chern number in TMD Moir\'{e} superlattice.}

\emph{Model.}--- We consider twisted homobilayer TMD materials. For each single layer, the low energy electronic states reside at the valence band maxima at $\pm \boldsymbol{K}$ valleys. Contrary to bilayer graphene systems where the valley and spin degrees of freedom are both present, in TMD  each valley in the top valence band has fixed spin orientation due to strong spin-orbit coupling and the broken inversion symmetry \cite{PhysRevLett.108.196802}. With a small twist angle $\theta$ between two layers, the $+\boldsymbol{K}$ valley for the top and bottom layers are shifted to $\boldsymbol{K}_t$ and $\boldsymbol{K}_b$ in the Moir{\'e} Brillouin zone (MBZ) respectively, [Fig. \ref{gapratio}(b)]. For convenience we choose the rhombus-shaped MBZ and set the point $\boldsymbol{M}=(\boldsymbol{K}_t+\boldsymbol{K}_b)/2$ as the origin. We employ the continuum model \cite{bistritzer_moire_2011} in which the Moir\'e Hamiltonian for the $+\boldsymbol{K}$ valley is:
\bea
H_+(\boldsymbol{k},\boldsymbol{r})=\left(\begin{array}{cc}-\frac{\hbar^{2}\left(\boldsymbol{k}-\boldsymbol{K}_{b}\right)^{2}}{2 m^{*}}+\Delta_{\mathfrak{b}}(\boldsymbol{r}) & \Delta_{T}(\boldsymbol{r}) \\ \Delta_{T}^{\dagger}(\boldsymbol{r}) & -\frac{\hbar^{2}\left(\boldsymbol{k}-\boldsymbol{K}_{t}\right)^{2}}{2 m^{*}}+\Delta_{\mathrm{t}}(\boldsymbol{r})\end{array}\right)
\label{Hp}
\eea
Here $m^*$ is the effective mass. The form of Moir\'e potential, $\Delta_{b,t,T}$, is dictated by the $D_3$ crystalline symmetry and a combination of $C_{2z}$ rotation followed by switching the two layers, and can be parameterized by~\cite{Wu2019}:
\bea
\Delta_{T}(\boldsymbol{r})&=&w\left(1+e^{-i \boldsymbol{G}_{2} \cdot \boldsymbol{r}}+e^{-i \boldsymbol{G}_{3} \cdot \boldsymbol{r}}\right) \nonumber\\
\Delta_{l}(\boldsymbol{r})&=&2 w_{z} \sum_{j=1,3,5} \cos \left(\boldsymbol{G}_{j} \cdot \boldsymbol{r}+l \psi\right),
\eea
where $l\in\{b, t\}=\{+1,-1\} $ and $\boldsymbol{G}_j$ is the Moir\'{e} reciprocal lattice vectors with length $|\boldsymbol{G}_j|=\frac{4\pi}{\sqrt{3}a_M}$ and polar angle $\frac{\pi (j-1)}{3}$. Here $a_M=a_0/\theta$ is the Moir\'e lattice constant for a small twisted angle $\theta$ and $a_0$ is the lattice parameter of TMD. The Hamiltonian for {the} valley $-\boldsymbol{K}$ can be obtained by {the} time-reversal symmetry $H_-(\boldsymbol{k},\boldsymbol{r})=H_+(-\boldsymbol{k},\boldsymbol{r})^*$. To be specific, we focus on twisted MoTe$_2$ homobilayer with typical parameters $( {\hbar^2}/{2m^* a_0^2},\ w_z,\ w,\ \psi )=(495   \ \mathrm{meV},\ 8\ \mathrm{meV},\ -8.5\ \mathrm{meV},\ -89.6^\circ  )$ \cite{Wu2019}.

The top valence band of {a TMD single layer} splits into multiple Moir\'{e} bands due to the Moir\'{e} potential.
As shown in Fig. \ref{gapratio}(c) and (d), when the twist angle is close to $\theta_0=1.38^\circ$, the top Moir\'{e} band becomes nearly flat. The flatness of a band can be characterized by a ratio of the gap between the nearest bands to its band width. For the top Moir\'{e} band, the ratio can be as large as $13$. When $\theta<3.1^\circ$, The top Moir\'{e} band is topological characterized by a {valley/spin} Chern number $C=\pm 1$ due to the skyrmion lattice pseudo spin textures of the Moir\'{e} potential \cite{Wu2019}. The Chern number for the opposite valley{/spin} is opposite as required by time-reversal symmetry.
{Thus, at the single-particle level, such TMD homobilayer realizes a quantum valley/spin Hall insulator.

We then introduce screened Coulomb interaction and project it to the nearly flat top Moir\'e band~\cite{Ashvin2019}:
\bea\label{eq3}
H_{\mathrm{int}}&=&\frac{1}{2A}\sum_{\boldsymbol{q}} : \rho(\boldsymbol{q})V(\boldsymbol{q})\rho(-\boldsymbol{q}) :  \nonumber\\
&=&\sum_{\boldsymbol{k},\boldsymbol{k}',\boldsymbol{q},\tau,\tau'}\frac{U}{2N_{\mathrm{cell} }}v(\boldsymbol{q})\lambda_{\tau,\boldsymbol{q}}(\boldsymbol{k})\lambda_{\tau',\boldsymbol{q}}(\boldsymbol{k}')^* \nonumber\\ &&C^\dagger_{\tau}(\boldsymbol{k})C^\dagger_{\tau'}(\boldsymbol{k}'+\boldsymbol{q})C_{\tau'}(\boldsymbol{k}')C_\tau(\boldsymbol{k}+\boldsymbol{q}),
\eea
where $\tau=\pm$ is the valley index, and $\lambda_{\tau,\boldsymbol{q}}(\boldsymbol{k})=\langle u_{\tau, \boldsymbol{k}}\ket{u_{\tau, \boldsymbol{k}+\boldsymbol{q}}}$ is the form factor originated from {the} projection. Here $v(\boldsymbol{q})={4\pi\tanh(qd)}/{\sqrt{3}qa_M}$ is the dimensionless screened Coulomb potential with $d$ the separation between the electrode and Moir\'e superlattice, {which is set to $d=2a_M$ in the calculations}. $A$ is system area and $N_{\mathrm{cell}}$ {is the number of the unit cells in the calculations}. The coefficient of $v(\boldsymbol{q})$ is chosen to make $U$ equal to the bare Coulomb potential between two particles separated by $a_M$. $C_\tau(\boldsymbol{k})$ is the annihilation operator for single particle state $\ket{u_{\tau,\boldsymbol{k}}}$. We neglect the weak intervalley impurity scattering process associated with a large momentum transfer, and therefore {the Hamiltonian} also {has a} valley $U(1)_v$ symmetry.
{In this model,} there are two competing symmetry breaking states: {an} intervalley coherent state that breaks the valley $U(1)_v$ symmetry and {an} valley{/spin} polarized state that breaks {the} time reversal symmetry. At half filling of the topmost band (account for valley degree of freedom), {our Hartree-Fock analysis and exact-diagonalization results both suggest that a valley-polarized state is energetically favored, which spontaneously breaks the time-reversal symmetry and leads to a interaction-induced Chern insulator~\cite{supplement}}.
{At fractional filling, in principle, two types of fractional topological states might emerge, a fractional Chern insulator or a fractional topological insulator~\cite{PhysRevLett.103.196803,stern_fractional_2016}, depending on whether the time-reversal symmetry is spontaneously broken or preserved \footnote{One may think of another possibility, analogous to the Halperin $(m_1,\ m_2,\ n_1)$ states. However, this state is not favored because of the opposite Chern number in the opposite valley. \cite{supplement}}, and our} exact diagonalization below show that the FCI is {favored and} stabilized in our system.

\begin{figure}
\includegraphics[width=3.4 in]{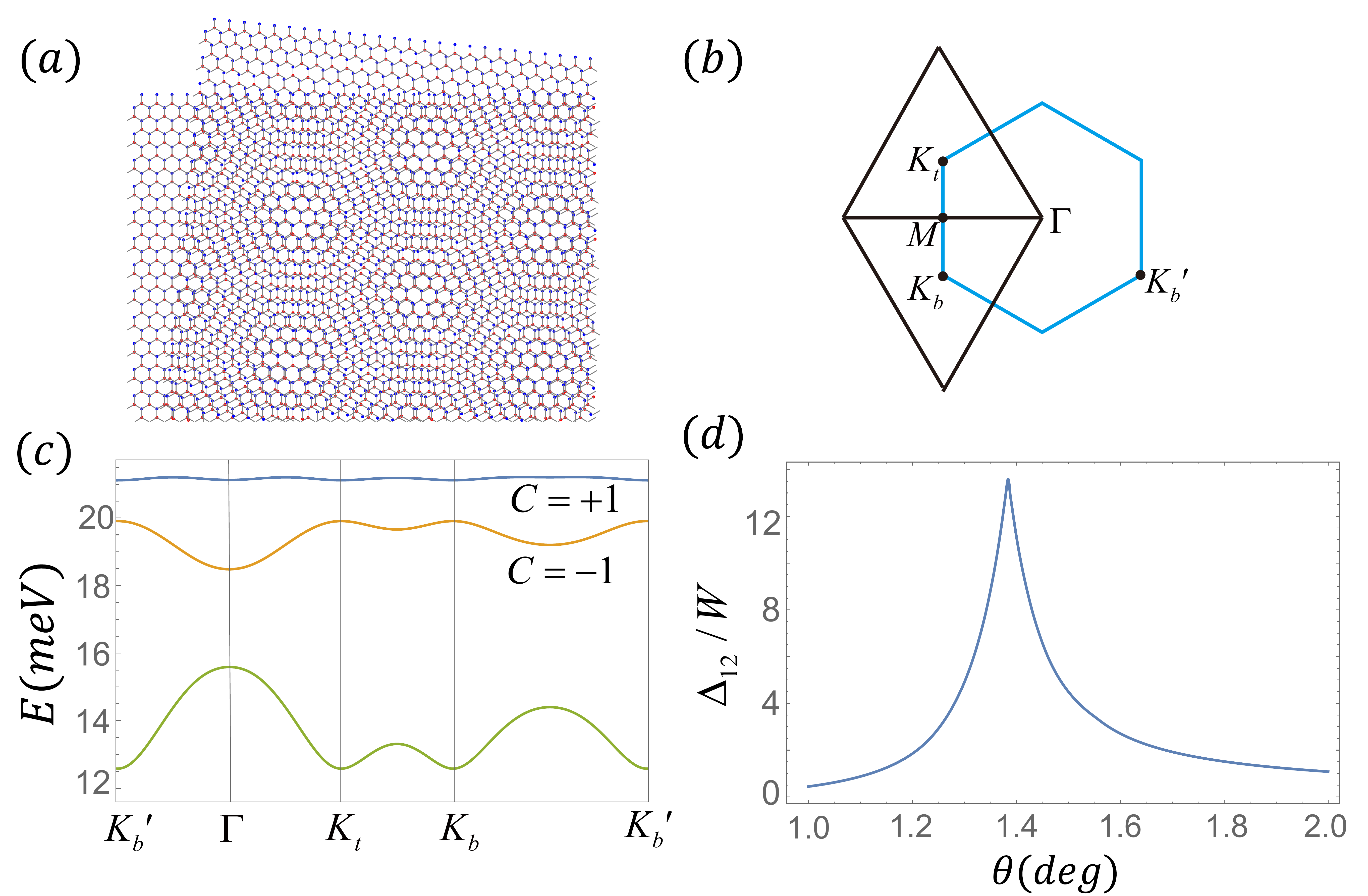}
\caption{(a): Schematic view of the Moir\'{e} superlattice. (b): We choose the Moir{\'e} Brillouin zone (MBZ) to be the rhombus and the origin in momentum space is chosen at $M$. (c): Moir{\'e} band structure at $\theta=1.38^\circ$. The top Moir{\'e} band is nearly flat with Chern number $\pm 1$. (d): The gap ratio $\frac{\Delta_{12}}{W}=\frac{\min(E_1(\boldsymbol{k}))-\max(E_2(\boldsymbol{k}))}{\max(E_1(\boldsymbol{k}))-\min(E_1(\boldsymbol{k}))}$ as a function of twisted angle $\theta$, where $E_1(\boldsymbol{k})$ ($E_2(\boldsymbol{k})$) is the energy of the first (second) topmost Moir{\'e} band.      }
\label{gapratio}
\end{figure}


\emph{Valley polarized FCI.}--- {We define the filling factor $\nu=2 \rho_e/\rho_s$, where $\rho_e$ is the electron density occupying the top Moir\'e band and $\rho_s$ is the electron density for the full filling of the two-fold degenerate top Moir\'e band. The factor 2 accounts for the valley degree of freedom.} Using exact diagonalization, at $\nu=1/3$ we observe numerical evidence of spontaneous valley polarization and FCI in the strong interaction limit, as shown in Fig. \ref{FCI}(a). {For 8 electrons in $4\times 6$ unit cells ($4\times 6\times 2$ single-particle states including both valleys)}, the ground states are fully valley polarized {with} three nearly-degenerate ground states {for each valley polarization}, separated from the excited states by an energy gap of the order of $2$ K.
{We calculated the many-body Chern number of each ground state~\cite{Sheng2011}, and the topological index is found to be $1/3$, characterizing a $1/3$ FCI phase. This conclusion is further supported by the total momentum for each ground state, which obeys the generalized Pauli exclusion rule~\cite{Bernevig2012}}.

The occupation number $n(k_1,k_2)$ of single particle states for each of the three many-body ground states are plotted in Fig. \ref{FCI}(b). $n(k_1,k_2)$ is uniformly distributed for different single particle states, consistent with the fact that the ground state is an incompressible liquid. The spectrum evolution under flux insertion along $k_2$ direction is shown in Fig. \ref{FCI}(c). The excitation gap is maintained throughout the flux insertion process.

\begin{figure}
\includegraphics[width=3.4 in]{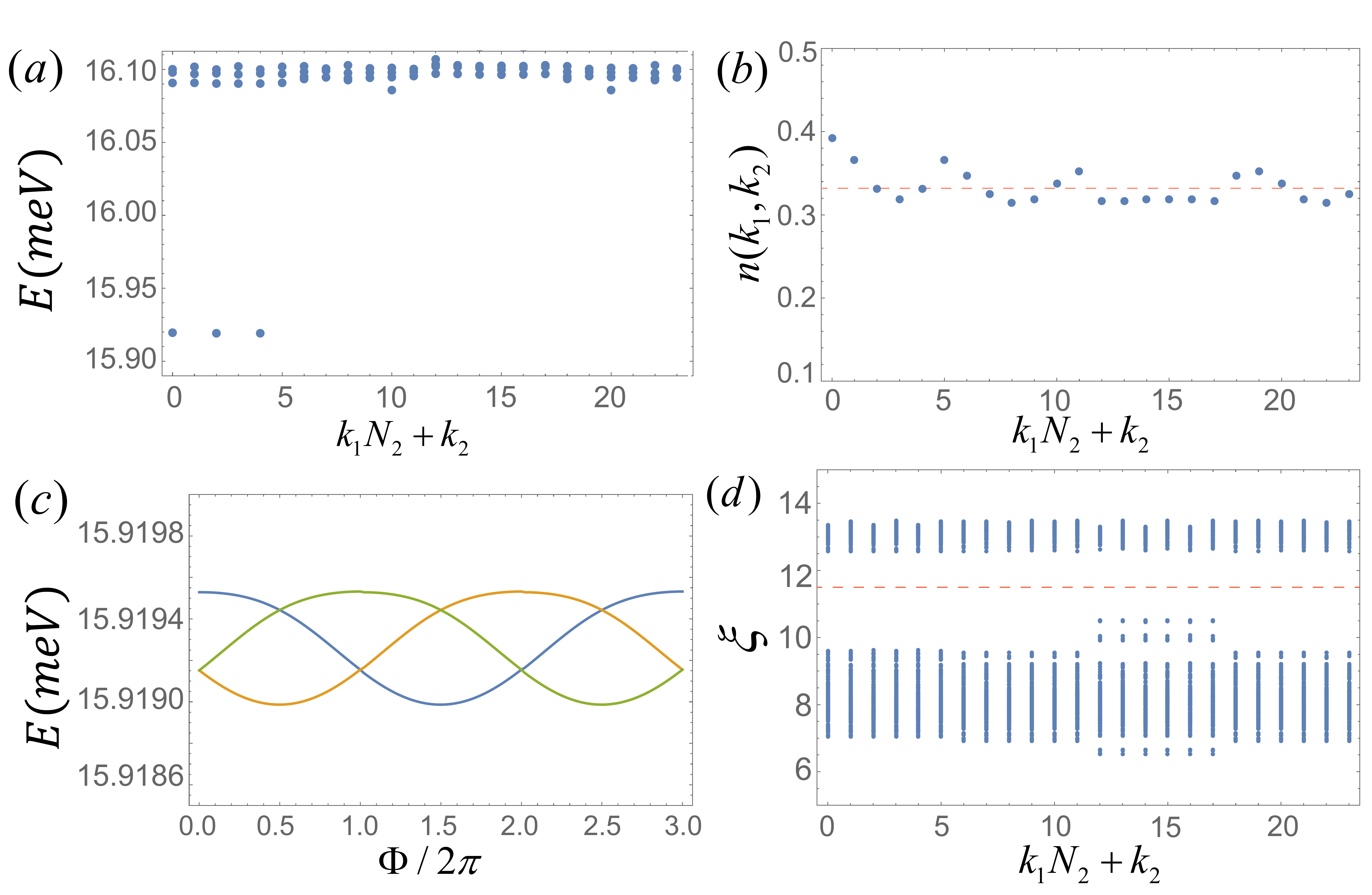}
\caption{Numerical diagonalization results for 8 particles in $4\times 6$ Moir{\'e} lattice. We choose $\theta=1.38^\circ$ and $U=1.38\ \mathrm{meV}$. The bandwidth at this twist angle is $W=0.083 \ \mathrm{meV}$. Here $N=N_1\times N_2/3$ is the number of particles. (a): Energy spectrum with three nearly degenerate ground states in each valley. (b): The occupation number of single particle states $n(k_1,k_2)$ for each of the three many-body ground states. The nearly uniform distribution of $n(k_1,k_2)$ suggests the ground state is an incompressible liquid. (c): Under flux insertion along $k_2$ direction, the ground states evolve into each other. (d): Particle entanglement spectrum (PES) for the separation of $N_A=4$ particles.}
\label{FCI}
\end{figure}

The topological nature of the ground states are further confirmed by our calculation of the particle entanglement spectrum (PES) \cite{Bernevig2012}.  To compute PES, we divide the $N$ particles into two collections of $N_A$ and $N_B=N-N_A$ particles and trace out $N_B$ particles to get the reduced density matrix $\rho_A$. The PES levels $\xi$ are obtained from the logarithm of eigenvalues of $\rho_A$, and are labeled by the total momentum of the remaining $N_A$ particles, as shown in Fig. \ref{FCI}(d). There is a clear entanglement gap with 2730 levels below the gap for $N_A=4$, consistent with the counting of quasihole excitation in the $\nu=1/3$ FCI \cite{Bernevig2012}.

To exam the finite-size effect, we study the scaling of the many-body gap $\Delta$ with various system sizes. For a genuine FCI, $\Delta$ remain finite in the thermodynamic limit when both $N_1$ and $N_2$ approach infinity. However $\Delta$ should vanish if only one of $N_1$ or $N_2$ approaches infinity, because this limit is a one-dimensional system which should not support FCI \cite{Regnault2011}. This is confirmed in Fig. \ref{Utheta}(a), which shows $\Delta$ decreases when $N_1$ is fixed at 3 and $N_2$ increases from 4 to 8, but $\Delta$ increases when the system size changes from $3\times 8$ to $4\times 6$.

\begin{figure}[b]
\includegraphics[width=3.4 in]{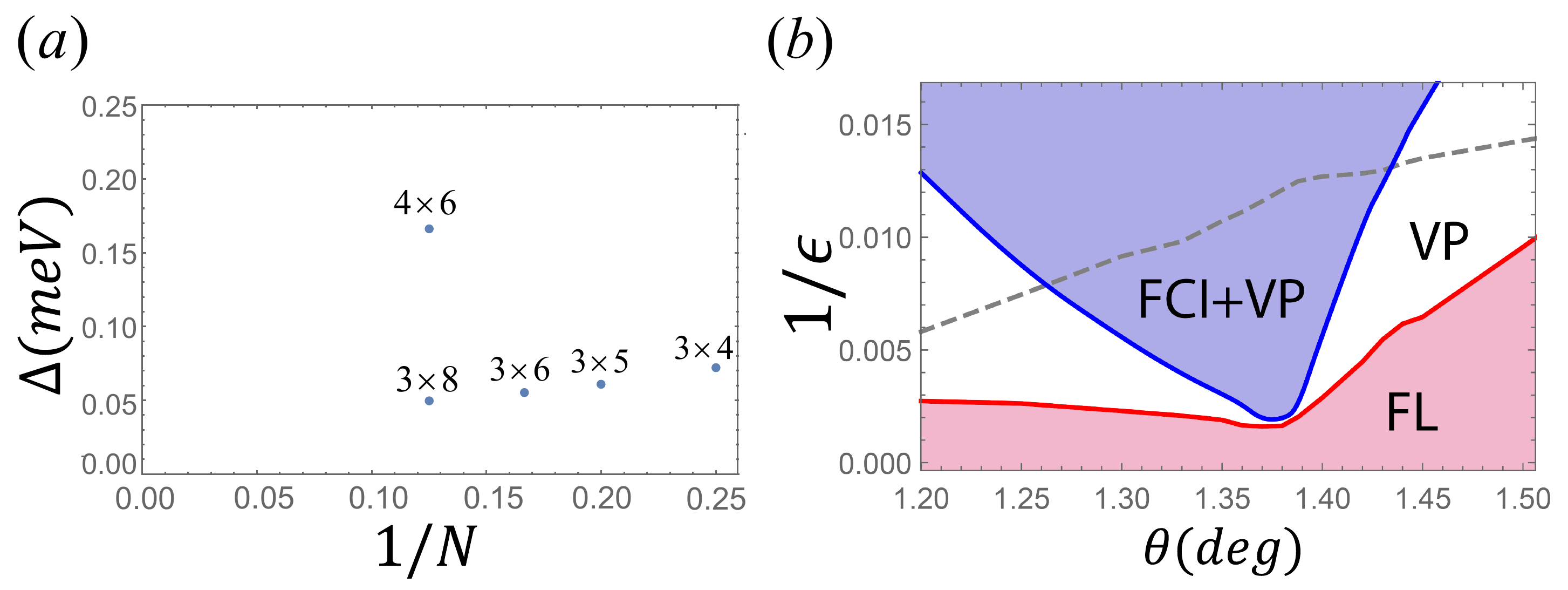}
\caption{(a): The many-body gap $\Delta$ for various system sizes at $v=1/3$ filling. The interaction strength is fixed to be $U=1.38\ \mathrm{meV}$. The increase of $\Delta$ in $4\times 6$ system suggests the gap persists in the two-dimensional thermodynamic limit. (b): The phase diagram for Fermi liquid (FL), FL with valley polarization (VP) and fractional Chern insulator (FCI) at different interaction strength $U(\epsilon)=\frac{e^2}{4\pi\epsilon\epsilon_0 a_M}$ and twisted angle $\theta$. The dashed line corresponds to $U(\epsilon)=\Delta_{12}$, above which the interaction starts to mix different bands and the single-band approximation breaks down.}
\label{Utheta}
\end{figure}

We then map out the phase diagram at $\nu=1/3$ filling as a function of the interaction strength $U$ which can be controlled by distance between the electrodes and {the} Moir\'{e} superlattice in experiments and the dielectric constant $\epsilon$. The results are shown in Fig. \ref{Utheta}(b). We find {a} valley non-polarized Fermi liquid at a small $1/\epsilon$ corresponding to a small $U$, a Fermi liquid with valley polarization at an intermediate interaction and the FCI phase with valley polarization at strong interaction. Depending on the twisted angle $\theta$, which controls the bandwidth, the valley non-polarized Fermi liquid can transit directly to FCI with valley polarization or through an intermediate Fermi liquid with valley polarization. The phase transition between the valley-polarized Fermi liquid and FCI can be described by Ginzburg-Landau theory with a Chern-Simons term \cite{supplement}. The direct transition occurs near $\theta=1.38^\circ$ where the single particle Moir\'e band has {the largest gap to bandwidth ratio} [see Fig. \ref{gapratio}(d)]. This is consistent with the quantum Hall systems with flat Landau levels, where the interaction stabilizes simultaneously the fractional quantum Hall state with spin polarization. Note that the FCI can be stabilized in a relatively broader region of the twisted angle here compared to that in magic angle twisted bilayer graphene \cite{PhysRevResearch.2.023237,PhysRevResearch.2.023238,PhysRevLett.124.106803,liu_gate-tunable_2020,wilhelm_interplay_2020,Sohal2018,Sohal2020}, and the region of angle for FCI increases with interaction. For interaction above the dashed line in Fig. \ref{Utheta}(b), our single-band approximation used in the numerical calculations breaks down and it requires to take other nearby bands into account.

\emph{Excitations.}--- Here we study the charge neutral excitations above the FCI ground states. As a consequence of the spontaneous valley polarization, we consider the valley waves excitation $\ket{\Psi_v(\boldsymbol{q})}=\sum_{\boldsymbol{k}} z_{\boldsymbol{k}} C_+^\dagger(\boldsymbol{k}+\boldsymbol{q}) C_-(\boldsymbol{k})\ket{\Psi_-}$, where $\ket{\Psi_-}$ is the FCI ground state with $\tau=-$ valley fully occupied and $z_{\boldsymbol{k}}$ is variational parameter. The presence of the form factor in Eq. \eqref{eq3} breaks the valley pseudospin $\mathrm{SU(2)}$ rotation system down to the valley $U(1)_v$ symmetry. As a result, the valley wave excitation are gapped as shown in Fig. \ref{valleywave}, which can be fitted by $E_w(\boldsymbol{q})=Jq^2+A$. The valley wave disperses weakly in momentum and thus is well localized in real space.

The lowest intravalley many-body excitation has lower energy than valley wave excitation for the parameters we used, i.e., the energy difference between the lowest fully-polarized excited state and the FCI state is $E_{mb}=0.167\ \mathrm{meV}<E_w$, see Fig. \ref{valleywave}. Nevertheless, the valley wave excitation remains a stable excitation because the decay of the valley wave to the intravalley many-body excitations are forbidden. Intravalley many-body excitations has valley quantum number 0, while the valley wave has valley quantum number 2.

In quantum Hall ferromagnets, a pair of skyrmions has lower energy than the particle-hole bound state \cite{PhysRevB.47.16419}.  The system size limitation in the exact diagonalization does not allow us to study the valley skyrmion excitation in our numerical calculations. Here we use an effective Hamiltonian density for valley pseudospin $\mathbf{n}(r)$ \cite{PhysRevB.47.16419}
\begin{align}
H_n(r)= \frac{J}{2} (\nabla \mathbf{n})^2-\frac{A}{2} n_z^2+\frac{1}{2}\int dr'^2 V(r-r')\rho_s(r)\rho_s(r'),
\end{align}
where $J$ and $A$ are given by the valley wave spectrum. The presence of the valley pseudospin anisotropy can be traced back to the opposite Chern number for the opposite valley. One cannot rotate $\mathbf{n}$ from one valley to opposite valley adiabatically without closing the energy gap, which implies the existence of anisotropy for $\mathbf{n}$. The last term accounts for the Coulomb interaction $V(r-r')$, because a skyrmion is dressed with charge distribution $\rho_s(r)=\epsilon^{ij}\epsilon_{abc}n^a\partial_i n^b\partial_j n^c/8\pi$, where $\epsilon_{abc} (\epsilon^{ij})$ is the Levi-Civita tensor with $i, j$ being the space index and $a, b, c$ being the spin index. The skyrmion topological charge $Q_s=\int d r^2 \rho_s(r)$ is quantized to an integer number. The easy axis anisotropy favors skyrmions with a small radius while the Coulomb repulsion favors skyrmions with a large radius. Their competition determines the skyrmion size \cite{chatterjee_skyrmion_2020}.

\begin{figure}
\includegraphics[width=2.8 in]{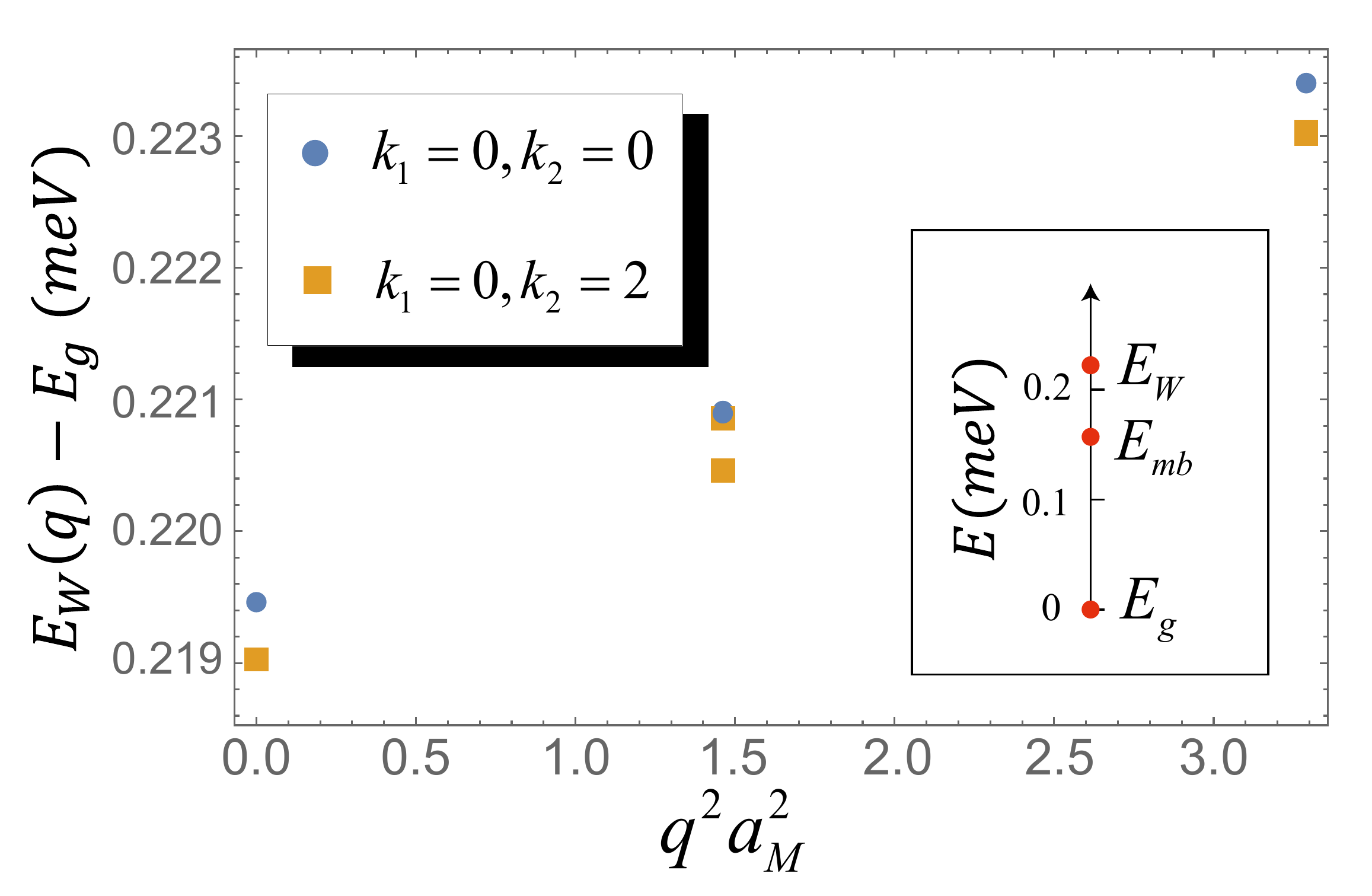}
\caption{Dispersion of valley wave excitation $E_w(\boldsymbol{q})$ for 8 particles in $4\times 6$ lattice. Excitation above the ground state with total momentum $k_1=0,k_2=0$ ($k_1=0,k_2=2$) are labeled by the squares and circles respectively. The slight energy difference for these two ground states is caused by finite size effect. The inset compares the energy of the lowest valley wave excitation $E_w$, the lowest intravalley many-body excitation $E_{mb}$ and the ground state energy $E_g$.}
\label{valleywave}
\end{figure}

\emph{FCI at $v=2/5$.}---  In twisted graphene Moir\'{e} superlattices, the Halperin $(332)$ state is stabilized at $v=2/5$ due to the remaining SU(2) spin rotation symmetry in the valley polarized state \cite{liu_gate-tunable_2020}. In our TMD Moir\'{e} superlattices, the spin rotation symmetry is absent because of the spin-valley locking. The Chern number contrasting valley degree of freedom disfavors the $(332)$ state. To demonstrate this explicitly, we calculate the energy spectrum, spectrum flow under flux insertion and entanglement spectrum at $v=2/5$, and the results are displayed in Fig. \ref{fig5}. The 5-fold degenerate ground states are valley polarized, and are consistent with the $v=2/5$ FCI state. In the fractional quantum Hall, the $v=2/5$ state belongs to the second hierarchical Jain state, and similarly one can assign the $v=2/5$ FCI as a second hierarchical FCI. Our results highlight the importance of symmetry in dictating the ground state and contrast the difference between the graphene and TMD Moire superlattices.

\begin{figure}[t]
\includegraphics[width=3.4 in]{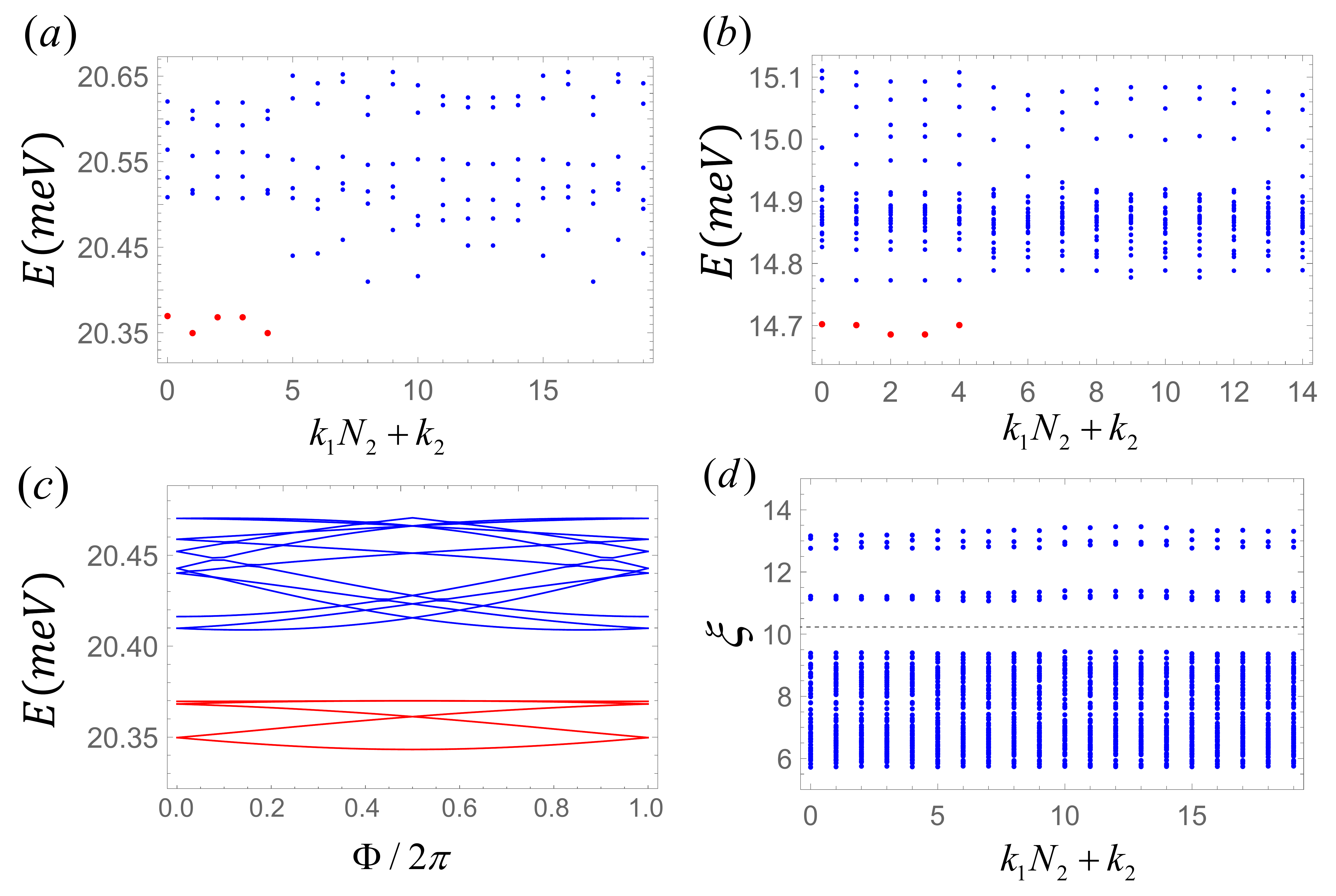}
\caption{ FCI at $v=2/5$. (a): the energy spectrum of 8 particles in 4*5 system, and (b): 6 particles in 3*5 system. (c): the flux insertion for the system in (a), where the five ground states are marked in red (some of them are on top of each other). A finite gap remains during flux insertion. (d): the particle entanglement spectrum for (a) with $N_A=3$. There are $51\times 20=1020$ states below the dashed line, consistent with quasihole counting.}
\label{fig5}
\end{figure}
}

\emph{Discussions.}--- We show that TMD Moir\'{e} superlattices can host fractional topological states via spontaneously breaking the time-reversal symmetry, using realistic parameters of TMD Moir\'{e} superlattices. Comparing with graphene, the spin-valley locking in TMD materials breaks the SU(2) spin rotation symmetry and eliminates the spin wave Goldstone modes, which could help stabilize the FCI states. The valley contrasting Chern number in TMD Moir\'{e} superlattices also dictates the symmetry breaking states, hence the nature of fractionalized topological states,  and also the low energy excitations in the FCI. The gapped nature of these states can be detected by transport, optical measurements etc, and its topological nature can be accessed by Hall conductivity measurement. Due to the strong analogy between FCI and chiral spin liquids, it is plausible that Moir\'{e} superlattices may also help realize/stabilize exotic spin liquid phases \cite{PhysRevB.90.174409,PhysRevB.92.094433,PhysRevLett.117.096402} by unitizing the valley/layer pseudospin or real spin degrees of freedom.

\begin{acknowledgements}

{\emph{Acknowledgements.}}--- This work done at LANL was carried out under the auspices of the U.S. DOE NNSA under contract No. 89233218CNA000001 through the LDRD Program.  S. Z. L. was also supported by the U.S. Department of Energy, Office of Science, Basic Energy Sciences, Materials Sciences and Engineering Division, Condensed Matter Theory Program. H.L. and K.S. acknowledge  support through NSF Grant No. NSF-EFMA-1741618.

\end{acknowledgements}


%

\widetext
\clearpage
\begin{center}
\textbf{\large Supplemental Material: Spontaneous fractional Chern insulators in transition-metal-dichalcogenides Moir{\'e} superlattices}
\end{center}
\setcounter{equation}{0}
\setcounter{figure}{0}
\setcounter{table}{0}
\makeatletter
\renewcommand{\theequation}{S\arabic{equation}}
\renewcommand{\thefigure}{S\arabic{figure}}
\renewcommand{\bibnumfmt}[1]{[S#1]}
\renewcommand{\citenumfont}[1]{S#1}


\subsection{I. Energy for valley polarized (VP) and intervalley coherent (IVC) state}
Here we compare the energy for the valley polarized and intervalley coherent state. We consider half filling of the topmost band including the valley degree of freedom. The wave function for the VP state can be written as
\[
\ket{\Psi _{\text{VP}}} =\prod _k  C_{+}^\dagger (k)\ket{0}.
\]
Here we choose $\tau=+$ valley to be fully occupied. Its energy $\bra{\Psi _{\text{VP}}}H\ket{\Psi _{\text{VP}}}$ can be decomposed into single particle contribution $E_0=\sum _k \epsilon _+(k)$; Hartree contribution: $E_{\mathrm{Ha}}=V(0)[\sum _k\lambda_{+, 0}(k)]^2$ and the Fock contribution: $E_{\mathrm{Fo}}=-\sum _{k, q}V(q)|\lambda_{+, q}(k)|^2$. Here the summation of $k$ is over the whole Moir\'{e} Brillouin zone.

We note that the system Hamiltonian is invariant under the following gauge transformation
\[
C_\tau(k)\rightarrow \exp[i\theta_\tau (k)]C_\tau(k),
\]
\[
\lambda_{\tau, q}(k)\rightarrow \exp[i\theta_\tau (k)-\theta_\tau (k+q)]\lambda_{\tau, q}(k).
\]
We choose a gauge which fixes the valley pseudospin associated with $\ket{\Psi _{\text{IVC}}}$ in the $x$ direction, and the wave function of the intervalley coherent state can be written as
\[
\ket{\Psi _{\text{IVC}}} =\frac{1}{\sqrt{2}}\prod _k  [C_{+}^\dagger (k)+C_{-}^\dagger (k)]\ket{0}.
\]
The energy for the IVC is the sum of single particle contribution $E_0'=\sum _k [\epsilon _+(k)+\epsilon _-(k)]/2$; Hartree contribution: $E_{\mathrm{Ha}}'=\frac{V(0)}{4}[\sum _{k,\tau=\pm}\lambda_{\tau, 0}(k)]^2$ and the Fock contribution
\[
E_{\mathrm{Fo}}'=-\frac{1}{4}\sum _{k, q}V(q)\left[|\lambda_{+, q}(k)|^2+|\lambda_{-, q}(k)|^2+\lambda_{+,q}(k)\lambda_{-,-q}(k+q)+\lambda_{-,q}(k)\lambda_{+,-q}(k+q)\right].
\]
It easy to check that the single particle and Hartree part of the energy for the VP and IVC are the same. We compare the Fock part of the energy. Noticing that $\lambda_{\tau,-q}(k+q)=\lambda_{\tau,q}^*(k)$, we have
\[
\lambda_{+,q}(k)\lambda_{-,-q}(k+q)+\lambda_{-,q}(k)\lambda_{+,-q}(k+q)\le 2|\lambda_{+,q}(k)\lambda_{-,-q}(k+q)|,
\]
where the bound is saturated when $\lambda_{+,q}(k)\lambda_{-,-q}(k+q)$ is positive real. The energy difference between the IVC and VP is
\[
E_{\mathrm{Fo}}'-E_{\mathrm{Fo}}\ge \frac{1}{4}\sum _{k, q}V(q)\left[|\lambda_{+, q}(k)|-|\lambda_{-, q}(k)|\right]^2.
\]
By considering time reversal oepration and $C_2$ rotation combined with layer flipping symmetry, we can show $|\lambda_{+, q}(k)|=|\lambda_{-, q}(k)|$. After we choose the form of $\ket{\Psi _{\text{IVC}}}$ by fixing a gauge, it is not guaranteed that $\lambda_{+,q}(k)\lambda_{-,-q}(k+q)$ is positive real for \emph{all} $k$ and $q$. Therefore the IVC always has higher energy than that of the VP state. This conclusion is further supported by more detailed Hartree-Fock calculations below.

\begin{figure}[h]
	\centering
	\includegraphics[width=0.80\linewidth]{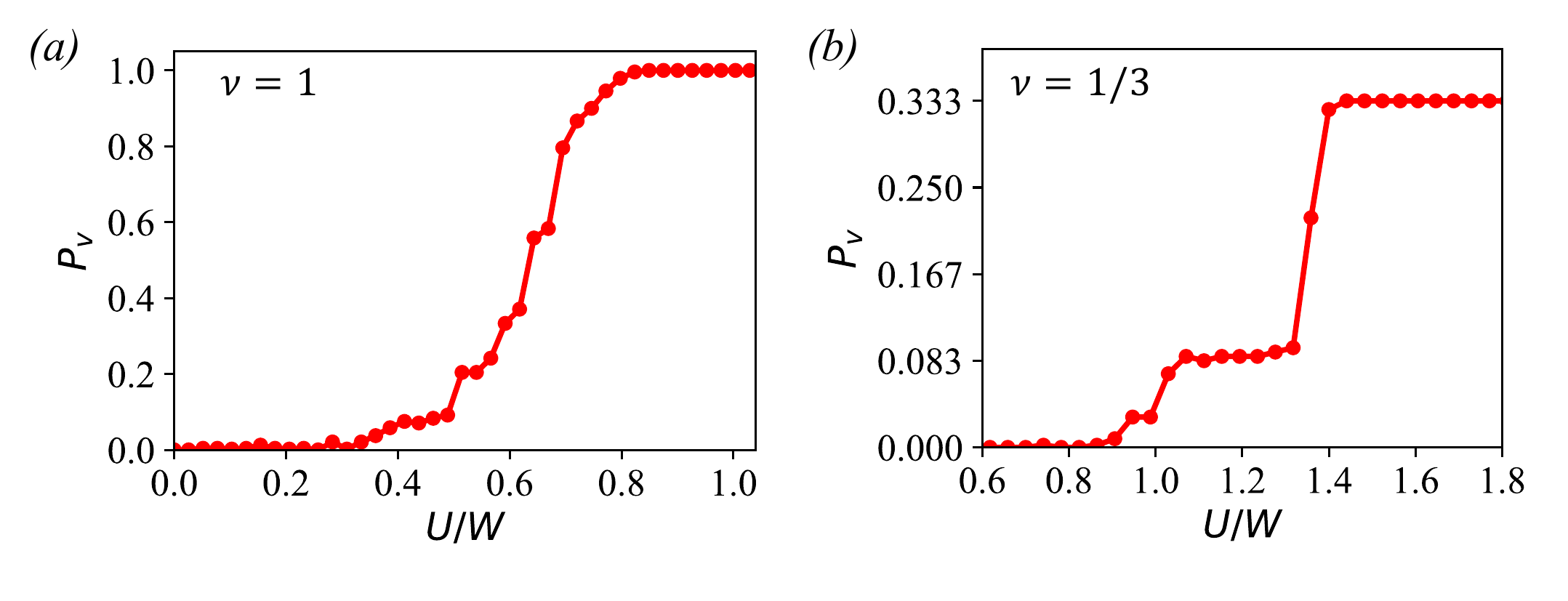}
	\vspace{-5mm}
	\caption{ $P_v~(= \sum_{\bm{k}}[\Delta_{++}(\bm{k}) - \Delta_{--}(\bm{k})])$ dependence on interaction ($U$) for  filling ($\nu=1$) in panel (a) and fractional filling ($\nu=1/3$) in panel (b). We observe unpolarized metal for smaller $U$, polarized metal for intermediate $U$ and a valley polarized state for large $U$.}
	\label{fig:pdintfracfilling}
\end{figure}

\subsection{II. Hartree-Fock calculations}
We treat the interaction part of the Hamiltonian using Hatree-Fock  mean-field  theory~\cite{PhysRevX.10.031034} in which the Hamiltonian can be written as
\begin{equation}\label{eq:MFHartreeFock}
\begin{split}
\mathcal{H}_{MF} = \sum_{k,\tau, \tau'} C_{\tau}^\dagger(\bm{k})[(h_0(\bm{k})-\mu)\delta_{\tau, \tau'}+h_{HF}^{\tau, \tau'}(\Delta_k,  \bm{k})]C_{\tau'}(\bm{k}) -\frac{1}{2} \text{tr}~h_{MF} (\Delta_k,  \bm{k}) \Delta_k^T
\end{split}
\end{equation}
Here, $h_0(\bm{k}) = h_{BM}(\bm{k})-\frac{1}{2}h_{HF}(\Delta_0,  \bm{k})$ where $\Delta_0$ is the reference density matrix such that  $\mathcal{H}_{MF} = h_{BM}$ of symmetry unbroken state when $\Delta_k =\Delta_0$~\cite{PhysRevX.10.031034}.  We therefore, choose $\Delta_0 = \begin{pmatrix}	\nu/2 & 0 \\ 	0& \nu/2 \end{pmatrix} ~\forall~\bm{k}$.

Also, the $h_{HF}$ is given by,

\begin{equation}\label{eq:Ham_HF}
\begin{split}
h_{HF}^{\tau, \tau'} (\Delta_k, k ) & = \frac{U}{2N_\text{cell}} \sum_{\bm{G}} v_{\bm{G}} \lambda_{\tau, \bm{G}}(\bm{k}) \delta_{\tau, \tau'}\sum_{\tau'', \bm{k}'} \text{tr}[ \lambda_{\tau'', \bm{G}}^\dagger (\bm{k}' ) \Delta_{\tau, \tau''}^{T}(\bm{k}) ]  \\
&-  \frac{U}{2N_\text{cell}} \sum_{\bm{G},\bm{k}'} v_{\bm{G}+\bm{k}'} \lambda_{\tau, \bm{G}+ \bm{k}'}(\bm{k}) \lambda_{\tau', \bm{G}+ \bm{k}'}^\dagger (\bm{k}) \Delta_{\tau, \tau'}^{T}(\bm{k}+ \bm{k}')
\end{split}
\end{equation}
In the above equation, the first and second terms are the Hartree and Fock contributions, respectively.
Here, $N_\text{cell}$ is the total area of the MBZ and we have used $\bm{q} = \bm{G}+ \bm{k}' $. $\bm{k}, \bm{k}'$ are momentum vectors in the first Brillouin zone (BZ) and $\bm{G}$ is the reciprocal vector connecting different BZs.  The matrix, $\lambda_{\tau, \bm{q}}(\bm{k})= \lambda_{\tau, \bm{G}+ \bm{k}'}(\bm{k}) $ contains the form factors for the single particle given by, $ \lambda_{\tau, \bm{G}+ \bm{k}'}(\bm{k}) = \langle u_{\tau,\bm{k}}|u_{\tau,\bm{k}+\bm{G}+\bm{k}'}\rangle $. We also have $v_{\bm{q}} = \frac{4\pi \tanh(qd)}{q \sqrt{3}a_M}$  for dual-gate screened Coulomb interaction.

For writing the mean-field equation, we use the following condition; a)  $\lambda_{+,\bm{q}} (\bm{k})= \lambda_{-,-\bm{q}}^* (-\bm{k})$, and b) the $\bm{G}_j^{th}$ component  of momentum,  $\bm{q} = \bm{G}_{l}+\bm{k}$   in the $l^{th}$ Moir{\'e} Lattice is generated using central BZs as, $|u_{\tau, \bm{k}+\bm{G}_l} (\bm{G}_j )\rangle = |u_{\tau, \bm{k}} (\bm{G}_j+\bm{G}_{l})\rangle$, so as to have a consistent gauge.

One can write the new quasiparticle by solving the above equation; {$V^\dagger H_{MF} V V^\dagger |\psi\rangle = E_n V^\dagger |\psi\rangle$. One can evaluate the Hamiltonian as $\langle \psi | V V^\dagger H_\text{MF} V V^\dagger |\psi\rangle = \langle \phi| D |\phi\rangle$}  where
\begin{equation}
\begin{split}
|\psi(\bm{k})\rangle=
\begin{pmatrix}
C_{+}(\bm{k})     \\
C_{-}(\bm{k})     \\
\end{pmatrix}
= V(\bm{k}) |\phi(\bm{k})\rangle
\end{split}
= \begin{pmatrix}
u_{1}(\bm{k})      &   u_{2}(\bm{k})       \\
v_{1}(\bm{k})     &  v_{2}(\bm{k})   \\

\end{pmatrix}
\begin{pmatrix}
\gamma_{1}(\bm{k})      \\
\gamma_{2} (\bm{k})      \\
\end{pmatrix}
\end{equation}
The gap $\Delta_{\tau, \tau'} ( \bm{k}) = \langle C_{\tau}^\dagger(\bm{k})  C_{\tau'}(\bm{k}) \rangle $ has to be written in terms of these new quasiparticles

We now write the gap equation in the new basis,
\begin{equation}\label{eq:MFOP}
	\Delta_{\tau, \tau'}(\bm{k}) = \langle C_\tau^\dagger (\bm{k}) C_{\tau'}(\bm{k})\rangle = \begin{pmatrix} |u_1(\bm{k})|^2 \langle n_{\gamma_1}(\bm{k}) \rangle +|u_2(\bm{k})|^2 \langle n_{\gamma_2}(\bm{k}) \rangle
		&  u_1^*(\bm{k}) v_1(\bm{k}) \langle n_{\gamma_1}(\bm{k}) \rangle +u_2^*(\bm{k})v_2 (\bm{k}) \langle n_{\gamma_2}(\bm{k}) \rangle \\
		u_1(\bm{k}) v_1^*(\bm{k}) \langle n_{\gamma_1}(\bm{k}) \rangle+u_2(\bm{k}) v_2 ^*(\bm{k}) \langle n_{\gamma_2}(\bm{k}) \rangle  &
		|v_1(\bm{k})|^2 \langle n_{\gamma_1}(\bm{k}) \rangle +|v_2(\bm{k})|^2 \langle n_{\gamma_2}(\bm{k}) \rangle \end{pmatrix}.
\end{equation}
Here $n_{\gamma_m}(\bm{k}) = \gamma_m^\dagger(\bm{k})\gamma_m(\bm{k})$
Also,  filling in the new basis is given by
\begin{equation}
\begin{split}\label{eq:filling}
\bar{n}& = \sum_{\tau, \bm{k}} \langle C_{\tau}^\dagger(\bm{k}) C_{\tau}(\bm{k})\rangle  =  \sum_{\bm{k}}  \text{tr}[ \Delta_{\tau, \tau'}]=\sum_{\bm{k}} \langle n_{\gamma_1}(\bm{k})  \rangle + \langle n_{\gamma_2}(\bm{k})  \rangle  = \nu
\end{split}
\end{equation}

Eqs.~\ref{eq:MFOP} and \ref{eq:filling} are then solved self-consistently for a fixed filling, until $\mu$ and mean field order parameter or all $\bm{k}$  converges. But, we use a relatively relaxed condition for convergence, as $\Delta_{\tau,\tau'} (\bm{k})$ can have multiple degenerate configurations, therefore, we use $\text{max}(\Delta^n - \Delta^{n+1} )< $ tolerance limit, where $\Delta  = \sum_{\bm{k}} \Delta_{\tau,\tau'} (\bm{k})$.   In the numerical simulation, we observe that only diagonal element of  $\Delta_{\tau, \tau'}(\bm{k})$ are populated whereas the off-diagonal elements are zero, {meaning that valley polarized state is the ground state. We define a valley polarization order parameter $P_v~= \sum_{\bm{k}}[\Delta_{++}(\bm{k}) - \Delta_{--}(\bm{k})]$.} \\

In Fig.~\ref{fig:pdintfracfilling}, we plot  $P_v$ dependence on the  interaction ($U$)  for $(\nu =1)$ and fractional $(\nu=1/3)$ fillings {using the parameters discussed in the main text and at $\theta = 1.38^\circ $}.
In the case of filling ($\nu =1$) shown in Fig.~\ref{fig:pdintfracfilling} (a), we observe that till around $U = 0.3 W$ we have unpolarized metal, $i.e.$ equal number of electron in both the valleys. For an intermediate interaction, $ 0.3W\le U \le  0.8W$, we observe partially polarized metal and from  $U = 0.8W$ onward the system is fully polarized. Here, $W$ is the bandwidth of the non-interacting bands. The band in one valley is fully occupied and the system is a valley polarized insulator. This behavior is consistent with the results reported for twisted bilayer graphene in the Ref.~\cite{PhysRevLett.124.166601}.

On the other hand, in the case of the fractional filling ($\nu =1/3$) shown in Fig.~\ref{fig:pdintfracfilling} (b), we observe the system to be unpolarized metal till $U = 0.85W$. In the $0.85W\le U \le 1.4W$ regime, partially polarized metal is observed and finally above,  $U = 1.4W$, the system saturates into a completely polarized metal. Note that in this filling fraction, one can only partially fill the lower mean-field band. Hence, the system remains a metal in contrast to the case with $v=1$.

\subsection{III. Possibility of the Halperin state}
Here we ague that it is unfavorable to host the Halperin $(m_1,\ m_2,\ n_1)$ states in TMD Moir\'{e} superlattice, because of the opposite Chern number in the opposite valleys. For a flat Chern band, a good starting point to understand the physics is the Landau level by neglecting the variation of the Berry curvature and band dispersion.  The Chern bands in the $\pm$ valleys in TMD Moir\'{e} superlattice can be treated as Landau levels stabilized by opposite effective magnetic field, $\pm B$. The wave function for the Halperin $(m_1,\ m_2,\ n_1)$ state is
\[
\ket{\Psi_H} =\prod _{i<j}\left(z_{+i}-z_{+j}\right)^{m_1}\left(z_{-i}-z_{-j}\right)^{m_2}\left(z_{+i}-z_{-j}\right)^{n_1} \exp \left[-\sum _{i,\tau =\pm }\frac{1}{4l_B^2}  | z_{\tau i}|^2\right],
\]
where $z_{\pm i}=x+i y$ and $l_B=\sqrt{\hbar c /e B}$ is  the magnetic length. Introducing composite particle through the flux attachment \cite{fradkin_field_2013}
\[
\phi_\tau(r) =\exp \left(i \Theta _{\tau }\right) \psi _{\tau }(r),
\]
\[
\Theta _+=m_1\int dr_2 \theta  (r_1-r_2) \rho _+ (r_2)+n_1\int  dr_2 \theta (r_1-r_2) \rho _- (r_2),
\]
\[
\Theta _-=m_2\int dr_2 \theta  (r_1-r_2) \rho _- (r_2)+n_1\int  dr_2 \theta  (r_1-r_2) \rho _+ (r_2),
\]
where $\theta  (r_1-r_2)$ is the angle between the vector $r_2-r_1$ and the $x$ axis. $\phi_\tau(r)$ is the composite particle wave function and $\psi_\tau (r)$ is the electron wave function. $\rho_\tau$ is the charge density. The effective magnetic experienced by the composite particles are
\[
B_{\text{eff},+}=B-\Phi _0 \left(m_1 \rho _++n_1 \rho _-\right),\ \ \ B_{\text{eff},-}=-B-\Phi _0 \left(m_2 \rho _-+n_1 \rho _+\right),
\]
with $\Phi_0=2\pi\hbar c/e$. It is not possible to make $B_{\text{eff},\pm}$ vanishing for any positive integers $(m_1,\ m_2,\ n_1)$. Therefore the Halperin state is generally unflavored in energetics.

\subsection{IV. Effective theory for the transition between the valley polarized Fermi liquid and fractional Chern insulator}
If we take the Landau level point of view by assuming completely flat bands with uniform Berry curvature, we can write down Ginzburg-Landau free energy to describe the phase transition between the Fermi liquid with valley polarization and fractional Chern insulator (FCI), analogous to the case for the transition between the fractional topological insulator and superfluid \cite{PhysRevB.85.195113}. In this consideration, we can think that two valleys experience an opposite magnetic field $B_\tau=\nabla\times a_\tau$. The Lagrangian can be written as
\[
\mathcal{L}_{\tau }= \bar{\psi } \left(i\partial_t -A_0-a_{\tau ,0}\right)\psi-\frac{1}{2m_\psi}| (-i \nabla -A-a_\tau)\psi|^2- V(\psi)+\frac{1}{4\pi m}\epsilon^{\mu\nu\lambda} a_\mu\partial_\nu a_\lambda,
\]
\[
\mathcal{L}_{\phi }= \bar{\phi }_\tau \left(i\partial_t -A_0\right)\phi_\tau-\frac{1}{2m_\phi}| (-i \nabla -A)\phi_\tau|^2- \alpha |\phi_\tau|^2-\frac{\beta}{2}|\phi_\tau|^4,
\]
\[
\mathcal{L}_{\phi }=-g |\phi|^2|\psi|^2.
\]
Here $a_\tau$ is the dynamical gauge field with Chern-Simons term. $A$ is the external electromagnetic gauge fields. $\psi$ describes the FCI condensate and $\phi_\tau=\langle C_\tau ^\dagger C_\tau\rangle$ is valley polarization order parameter, and their spatial average value obey $\langle|\psi|^2+|\phi_\tau|^2\rangle=\rho_0$ with $\rho_0$ the electron density. Here $m$ is an integer and we consider electron filling at $1/m$. The term with $g>0$ describes the repulsion between the FCI condensate and valley polarization condensate.  $V(\psi)$ is the potential for the FCI condensate which depends on the electron interaction.

From this construction, it is {clear} why the Fermi liquid with valley polarization (VP) is favored over the the Fermi liquid with inter-valley coherent state (IVC). In VP, the effective magnetic field due to the valley contrasting Chern band cancels. In $\mathcal{L}_{\phi }$, there is no $a_\tau$ gauge field, which is energetically favorable because of energy cost associated with the Meissner screening of $a_\tau$ if it is present. Whereas for the IVC, we need to condense $\phi_{\mathrm{IVC}}\sim \langle C_+^\dagger C_-\rangle $. In this case, we will have coupling to the $a_\tau$ field:  $-\frac{1}{2m_{\mathrm{IVC}}}| (-i \nabla -A-2 a_\tau)\phi_{\mathrm{IVC}}|^2$, which costs energy due to the Meissner effect.

The transition from the Fermi liquid with VP to FCI can be understood as follows. For an intermediate interaction, the Fermi liquid with VP is favored. Due to the repulsion between the FCI and VP condensates, electrons condense into the $\phi_\tau$ channel and the Fermi liquid with VP is stabilized. When the interaction becomes strong, the FCI is favored by the $V(\psi)$ term, and more electrons condense into the FCI state.
\end{document}